\documentclass[12pt,preprint]{aastex}

\newcommand{\myemail}{enomoto@icrr.u-tokyo.ac.jp}

\slugcomment{To appear in ApJ v596n1, Oct. 10, 2003.}

\shorttitle{Constraints on CDM in NGC 253}
\shortauthors{Enomoto, Yoshida, Yanagita, Itoh}

\begin{document}

\title{Constraints on Cold Dark Matter in the Gamma-ray Halo of NGC 253}

\author{
R.~Enomoto\altaffilmark{1},
T.~Yoshida\altaffilmark{2},
S.~Yanagita\altaffilmark{2},
and C.~Itoh\altaffilmark{3}
}

\altaffiltext{1}{Institute for Cosmic Ray Research, University of Tokyo,
Chiba 277-8582, Japan}
\altaffiltext{2}{Faculty of Science, Ibaraki University, Ibaraki 310-8512, Japan}
\altaffiltext{3}{Ibaraki Prefectural University of Health Sciences, Ibaraki 300-0394, Japan}

\email{\myemail}

\begin{abstract}
A gamma-ray halo in a nearby starburst galaxy NGC 253 was found by
the CANGAROO-II Imaging Atmospheric Cherenkov Telescope (IACT).
By fitting the energy spectrum with expected curves from
Cold Dark Matter (CDM) annihilations,
we constrain the CDM-annihilation rate
in the halo of NGC 253.
Upper limits for the CDM density were obtained in the wide mass
range between 0.5 and 50 TeV.
Although these limits are higher than the
expected values, it is complementary important
to the other experimental techniques, 
especially considering the energy coverage. 
We also investigate the next astronomical targets to improve 
these limits.
\end{abstract}


\section{Introduction} 

Recently, the gamma-ray halo in the nearby starburst galaxy NGC 253
was detected by CANGAROO-II \citep{itoh1,itoh3}.
Although it seems to be successful to interpret it as a high-energy
cosmic-ray halo \citep{itoh2},
we assumed that this radiation is due 
to cold dark matter (CDM) annihilation,
and obtained the upper limits of the CDM
density in a wide mass range around TeV, 

The motivation of this study is the morphology obtained in a
TeV gamma-ray observation \citep{itoh1,itoh3}. It marginally differed from a
disk shape. Considering the existence of this halo \citep{ostriker},
it would be worth obtaining the constraint of CDM
by using the observed TeV emissions.

In this paper we assumed two processes: 
inclusive gamma-ray production via
the annihilation of weakly interacting
massive particles to
quark anti-quark pairs \citep{rudaz}, 
and 
monochromatic gamma-rays
from the annihilation to two gamma state \citep{bergstrom}.
The former final states 
produce highly multiple
gamma-rays and gave better upper limits than the latter method. 
The final-state gamma-rays would show
an exponential energy spectrum which differs 
from the usually known cosmic-ray spectrum,
i.e., a power law.

\section{Property of NGC 253}

The nearby starburst galaxy NGC 253 is located inside the Sculptor group,
and can be clearly seen from the southern hemisphere. The distance was
estimated to be 2.5 Mpc \citep{vau}.
It was classified as SABc (Hubble classification), and is one of
the closest examples of ``our Galaxy-like object".
It is ``edge-on", i.e., suitable to distinguish its halo from the disk. 
The optical \citep{beck} and radio halos \citep{hummel,carilli}
 were previously observed in this galaxy, the sizes of which ($\sim$
10 kpc) approximately
agreed with a TeV-gamma-ray observation (13 - 26 kpc) \citep{itoh3}.

HI studies on the Sculptor group galaxies were carried out \citep{puche}.
Calculating the rotation curves, they concluded that many galaxies in this
group have massive halos.
Especially, that of NGC 253 was estimated to have a density of
0.015 M$_{\odot}$pc$^{-3}$ with a radius of 26.9 kpc, which is also
within the size estimation of the TeV-gamma-ray halo \citep{itoh3}. 

\section{Energy spectrum and mass of CDM}

\begin{figure}[htb]
\plotone{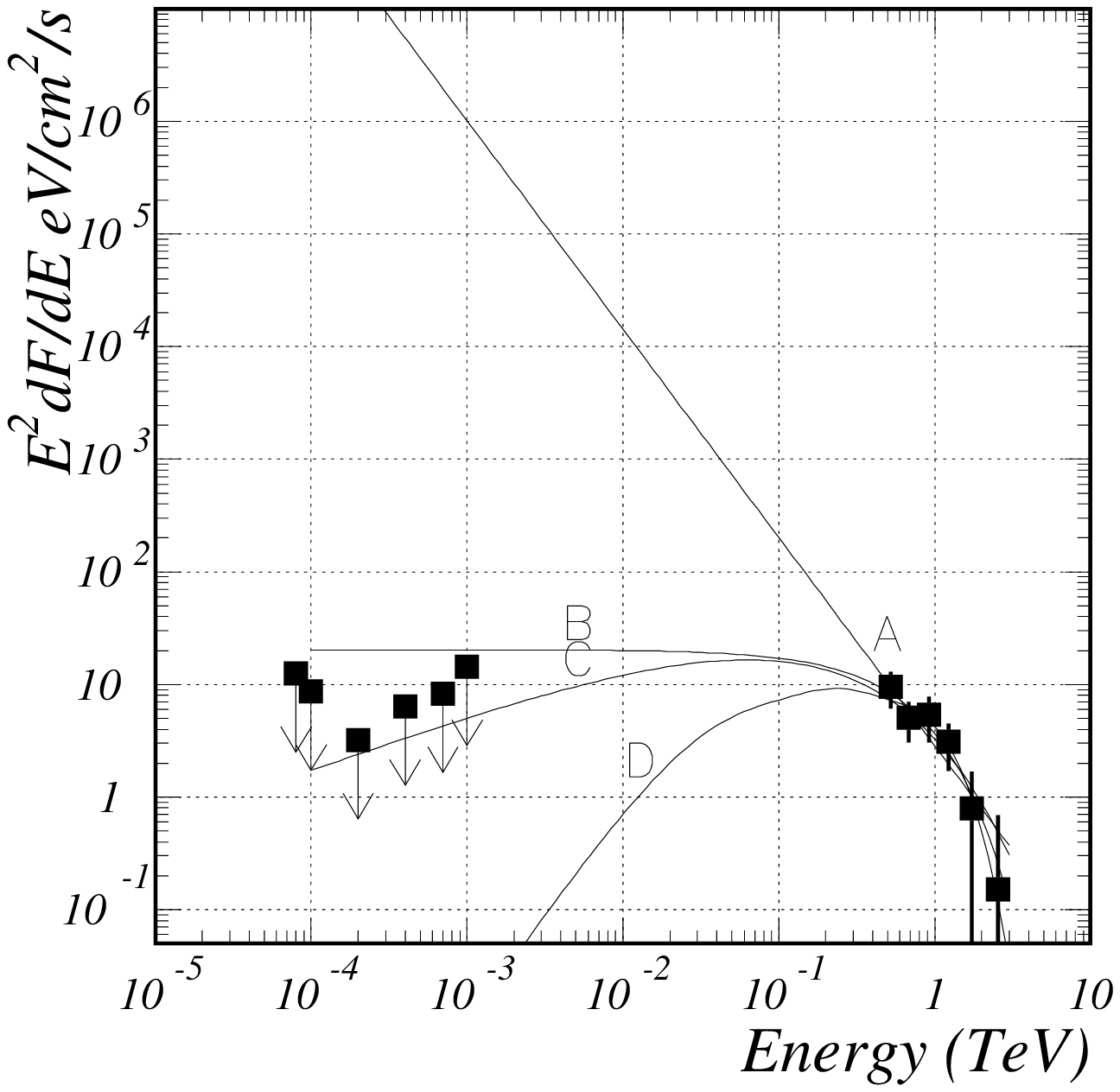}
\figcaption{
Spectral energy distribution. The high-energy data were obtained
by CANGAROO-II and the low-energy upper limits by EGRET.
The lines are the results of various fitting functions describe in the
text.
\label{fig1}}
\end{figure}

The spectral energy distribution (SED) of GeV-TeV gamma 
rays is plotted in Fig. \ref{fig1}.
The points with error bars were obtained by CANGAROO-II (Table
6 of \citet{itoh3}).
The arrows are upper limits obtained by EGRET \citep{egret1,egret2}.
According to a theoretical estimation motivated by the cosmic-ray 
radiation \citep{itoh2}, neither a simple power-law spectrum
($\propto E^{-\gamma}$) of
cosmic-rays (curve-A) nor that with a high energy cutoff
($\propto E^{-2}e^{E/E_{max}}$) (curve-B:
due to
$\pi^0\rightarrow \gamma \gamma$ decays or bremsstrahlung)
could simultaneously explain both data.
These curves can be well fitted to the TeV data, however,
they are inconsistent with the GeV upper limits.
The only choice
to satisfy both data was inverse Compton scattering oriented by very hard
incident electrons
($\propto E^{-1.5}e^{\sqrt{E}/b}$) (curve-C), 
which may require another mechanism
of re-acceleration in the galactic halo.

\begin{figure}[htb]
\plotone{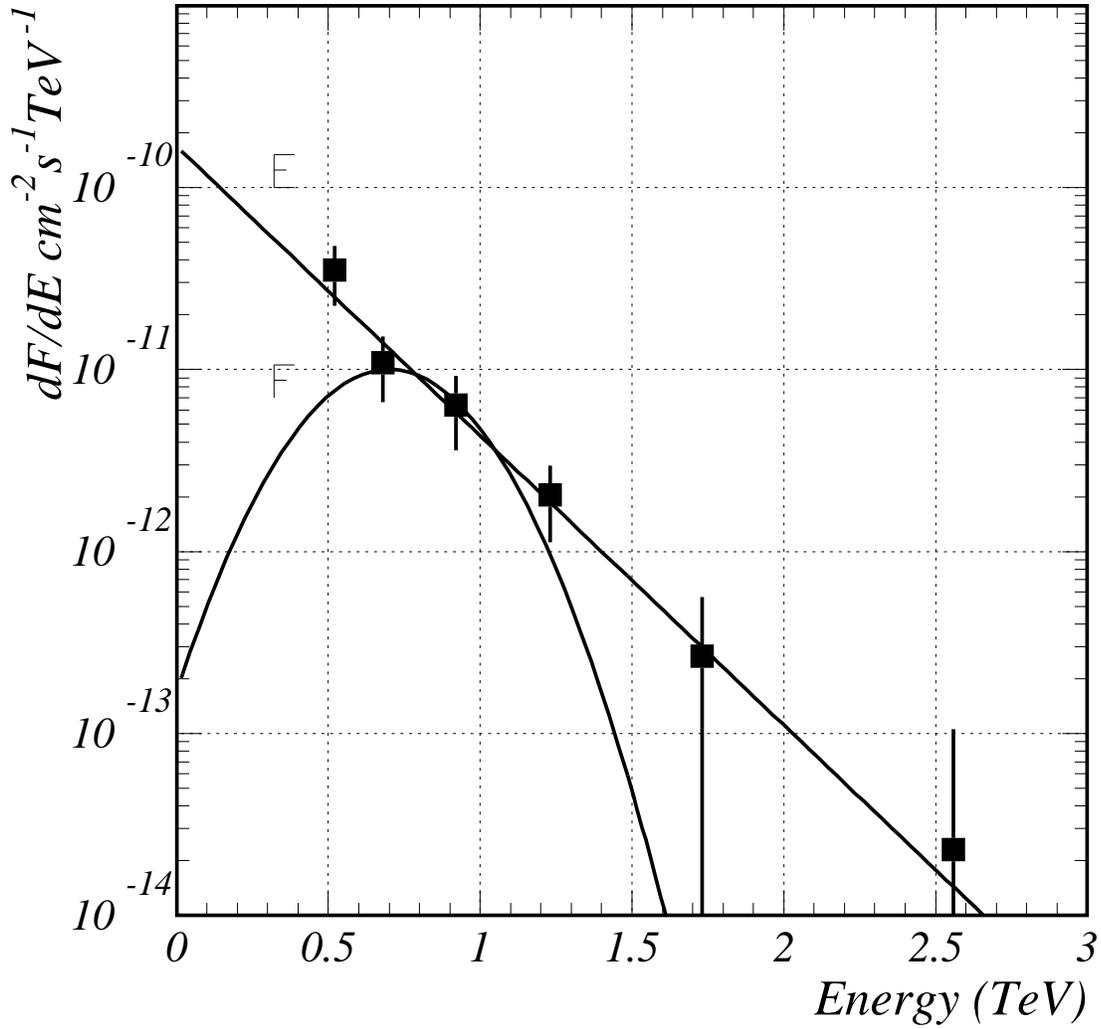}
\figcaption{
Differential flux of gamma-rays from NGC 253 in the semi-log scale. 
The data were obtained
from CANGAROO-II. Line-E is the best-fitted exponential curve and
line-F is an example of a Gaussian with an energy resolution of 35\%
and a center value of 0.7 TeV.
\label{fig2}
}
\end{figure}

Fig. \ref{fig2} shows 
a semi-log plot of the differential 
flux of TeV gamma-rays.
The line-E is the best-fitted exponential function
($\propto e^{-aE}$), and shows an agreement.
The extrapolation to GeV region is well below the EGRET upper limits.

The exponential function has a physical scale (in this case energy scale)
and its contribution in the GeV region is negligibly small in SED.
The well-known physical process to obtain an energy scale is a fragmentation
function ($\frac{1}{\sigma _h} \cdot \frac{d\sigma}{dx}$, 
where $x$ is a Feynman $x$) 
for such an inclusive particle spectrum as 
$e^+e^-\to q\bar{q}\to\gamma X$. It is typically to be fitted with
the sum of the exponential functions. The fragmentation function of
LEP data ($e^+e^-$ collider experiment at the center of mass energy of
$\sim$90 GeV) \citep{lep} was well-fitted with 
the sum of three exponential functions:
\begin{eqnarray*}
\frac{1}{\sigma _h} \cdot \frac{d\sigma}{dx} 
&=& e^{5.5605-34.482x} + e^{3.1777-10.551x}\\
& &+ e^{7.2391-123.29x}.
\end{eqnarray*}
Introducing the energy scale $M_\chi$, 
the relationship between $x$ and the energy of gamma-ray becomes
$x=E/M_\chi$. The annihilation rate ($F$ [cm$^{-2}$s$^{-1}$]) 
and the energy scale
were obtained by fitting the TeV-gamma-ray's spectrum with
$\frac{F}{\sigma _h} \cdot \frac{d\sigma}{M_\chi \cdot dx}$ to be
\begin{eqnarray*}
F&=&(1.8\pm 1.1) \times 10^{-11}\ [{\rm cm}^{-2}{\rm s}^{-1}],\\ 
M_\chi &=& (3.0\pm 0.6)\ [{\rm TeV}], 
\end{eqnarray*}
where we used the TeV gamma-ray fluxes from Table 6 in Ref. \citep{itoh3}.
Note that these two parameters are highly anti-correlated, which 
will be reflected in further analysis,
i.e., described later.
Here, $F$ is the observed annihilation rate 
per unit area and time at Earth.
The result is shown by line-D in Fig. \ref{fig1}.
The $\chi^2$ obtained in this fitting was 1.0/D.O.F=4.
The EGRET upper limits are also cleared.

For the reaction of CDM to $\gamma \gamma$ (i.e., $\chi \chi \rightarrow
\gamma \gamma$), the monochromatic gamma-rays were suggested to be searched 
\citep{bergstrom,bouquet,jungman}. The energy resolution of TeV gamma-rays
is approximately 35\% (Table 5 of \citet{itoh3}). 
The curve-F in Fig. \ref{fig2} is an example of a Gaussian 
($\propto e^{-\frac{1}{2}\left(\frac{E-M_\chi}{\sigma_E}\right)^2}$)
with that resolution and a center value of 0.7 TeV.

\section{Upper limit for the number density of CDM}

The annihilation in the volume of the halo should
be detected at a rate of
$$F=<\sigma v>B_{q\bar{q}}n^2[V/(4\pi d^2)],$$
where $\sigma$ is the annihilation cross section,
$v$ the relative velocity of CDMs, 
$B_{q\bar{q}}$ the branching fraction of $\chi\chi\to q\bar{q}$, 
$n$ the number density of CDM, $V$ the total volume of the halo,
and $d$ the distance from Earth.

Here, we consider the last volume-distance factor. The dark halo size
of NGC 253
obtained by an HI measurement is 26.9 kpc \citep{puche}, which
corresponds to a solid angle of $\Delta \theta = 0.62^o$.
Thus the volume factor becomes $\frac{d}{3}\cdot(\Delta \theta)^3$,
proportional to the distance. When we see the same-angular-sized diffuse image,
it suggests that distant objects have advantages. For example,
comparing the Galactic Center (distance of 8.5 kpc) 
and NGC 253, this factor becomes 300.
Although the Galactic Center may have a CDM concentration factor of, say,
1000 \citep{navarro}, it is highly model dependent.
On the other hand, the total volume average of the squared density
is less model dependent (only a factor changes 
under the assumption of $r^{-n}$, n=0,1,..).

The annihilation cross section is another source of model dependences.
For example, whether CDM is Dirac or Majorana fermion. Also,
it depends on the details of particle physics,
i.e., the details of SUSY breaking \citep{jungman2}.
A much larger dependence is expected to the $\chi\chi\to\gamma\gamma$
process. In order to avoid it, we carried out the following.
According to the Lee-Weinberg equation of the CDM density evolution
 \citep{lee,jungman2}, the annihilation cross section is 
directly related
to the cosmological abundance of $\Omega_{CDM}$,
$$\Omega_{CDM}=7\times 10^{-27}\ [{\rm cm}^{3}{\rm s}^{-1}]/<\sigma v>,$$
which is mass independent.
Recently, WMAP determined $\Omega_{CDM}=0.23$ \citep{wmap}.
With this, we normalized $<\sigma vB_{q\bar{q}}>$ to
an order of $10^{-26}\ [{\rm cm}^{3}{\rm s}^{-1}]$.

\begin{figure}[htb]
\plotone{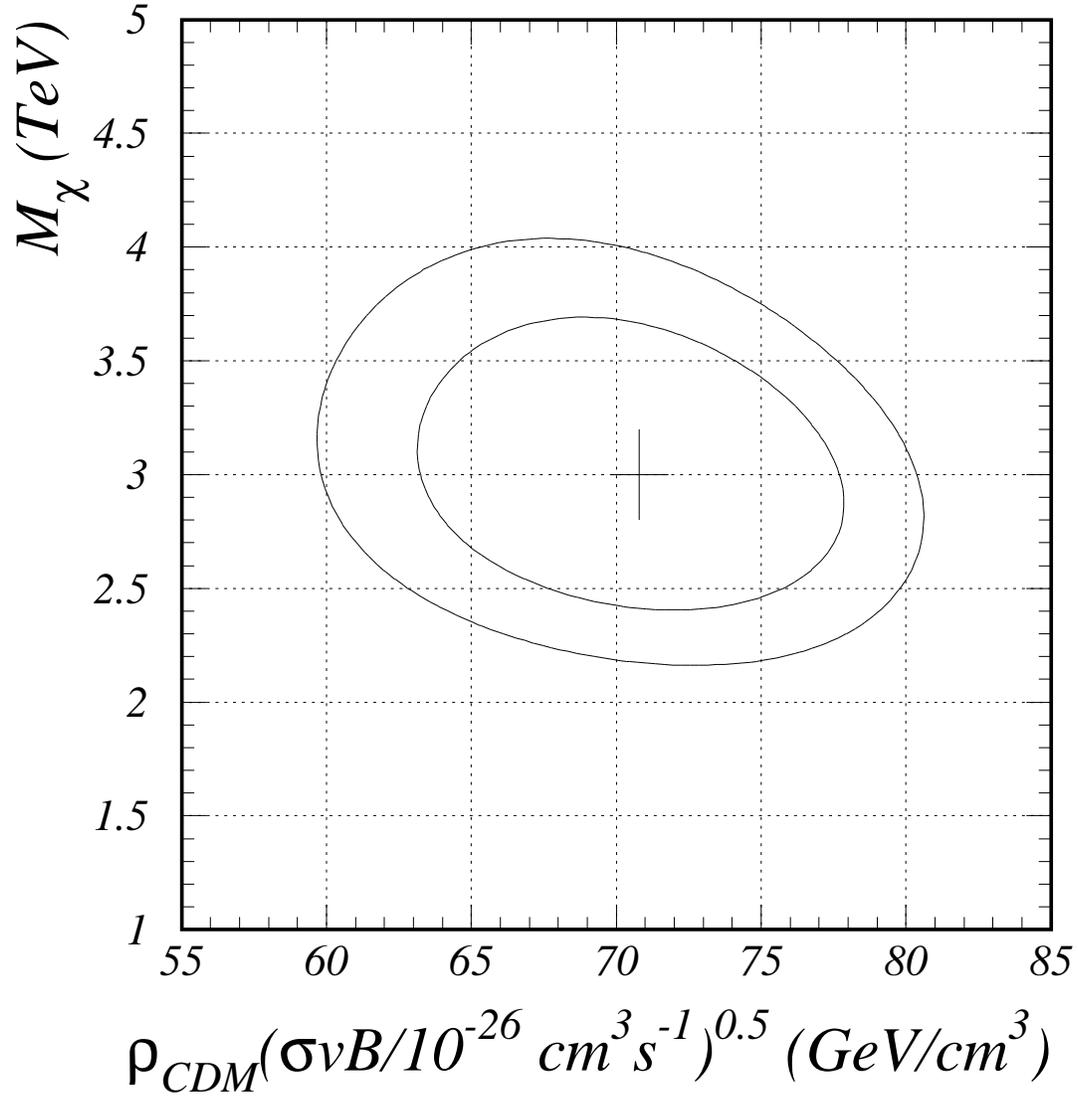}
\figcaption{
Correlation between $M_\chi$ and $\rho_{CDM}$. 1- and 2-$\sigma$
contours are shown with the cross corresponding to the best-fitted value.
\label{fig3}
}
\end{figure}

Now that all of unknown factors have been filled, 
by fitting the TeV gamma-rays spectrum with the described function,
we can derive the best-fitted density for CDM,
\begin{eqnarray*}
n=(2.4 \pm 0.6) \times 10^{-2}\sqrt{\frac{
10^{-26} {\rm cm}^{3}{\rm s}^{-1}}{<\sigma v B_{q\bar{q}}>}}[{\rm cm}^{-3}],
\end{eqnarray*}
where $n$ is still highly correlated with the $M_\chi$ value.
Changing $n$ to the energy density of the CDM,
\begin{eqnarray*}
\rho_{CDM}=(70.8 \pm 7.4) \sqrt{\frac{
10^{-26} {\rm cm}^{3}{\rm s}^{-1}}{<\sigma v B_{q\bar{q}}>}}
[{\rm GeV}{\rm cm}^{-3}]
\end{eqnarray*}
is obtained, where the correlation between $M_\chi$ and $\rho_{CDM}$
is shown in Fig. \ref{fig3}.
Note that $n$ and $\rho_{CDM}$ are the root-mean-squared
volume average of those densities.
The lines are 1- and 2-$\sigma$ contours.
The reasons why the errors became smaller compared to the value of $F$
are because $\rho_{CDM}$ is proportional to $\sqrt{F}$ and
there is a strong anti-correlation between $F$ and $M_\chi$.

To summarize, the final fitting data and functions are as follows:
\begin{eqnarray*}
\left[\frac{dF}{dE}\right]&=&\frac{F}{M_\chi}\cdot \left[\frac{1}{\sigma _h} 
\cdot \frac{d\sigma}{d(E/M_\chi)}\right]\\ 
 &=& \frac{<\sigma vB_{q\bar{q}}>n^2[V/(4\pi d^2)]}{M_\chi}\\
 & & \hskip 3cm\cdot \left[\frac{1}{\sigma _h}
\cdot \frac{d\sigma}{d(E/M_\chi)}\right] \\
 &=& \frac{<\sigma vB_{q\bar{q}}>\rho_{CDM}^2[V/(4\pi d^2)]}{M_\chi^3}\\
 & & \hskip 3cm\cdot \left[\frac{1}{\sigma _h}
\cdot \frac{d\sigma}{d(E/M_\chi)}\right], 
\end{eqnarray*}
where $[\frac{1}{\sigma _h} \cdot \frac{d\sigma}{d(E/M_\chi)}]$ 
should be replaced
with the linear combination of three exponential functions 
described so far and $[\frac{dF}{dE}]$
are Table 6 of \cite{itoh3} and Fig. 1 of \cite{egret2}, respectively.

\begin{figure}[htb]
\plotone{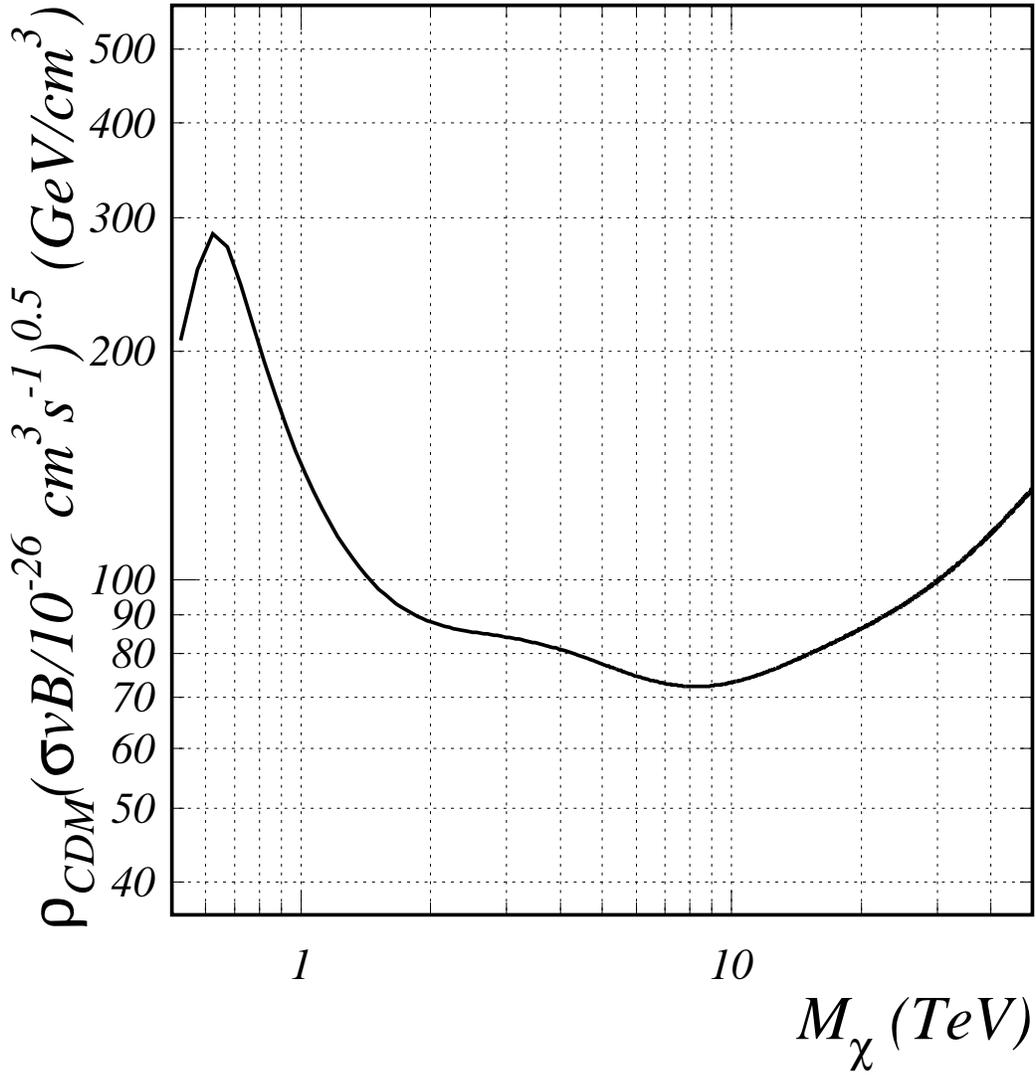}
\figcaption{
2$\sigma$-Upper limits of $\rho_{CDM}$ versus $M_\chi$ 
for various $M_\chi$ assumptions.
\label{fig4}
}
\end{figure}

Although, these values are the best-fitted ones,
there are not enough reasons to insist that this is evidence for CDM 
other than that
the energy spectrum was fitted well with the exponential
function. 
This value should be considered to be an upper limit.
We, therefore, carried out a scan for various $M_\chi$ assumptions,
and obtained upper limits versus $M_\chi$.
The results are shown in Fig. \ref{fig4}. 
Considering the dynamic range of the fragmentation function measurements,
the searched range was selected to be from 0.5 to 50 TeV.
In the figure, the 2$\sigma$-upper limits are shown. 
The improvement bellow 0.65 GeV was due to the EGRET upper limits.

\begin{figure}[htb]
\plotone{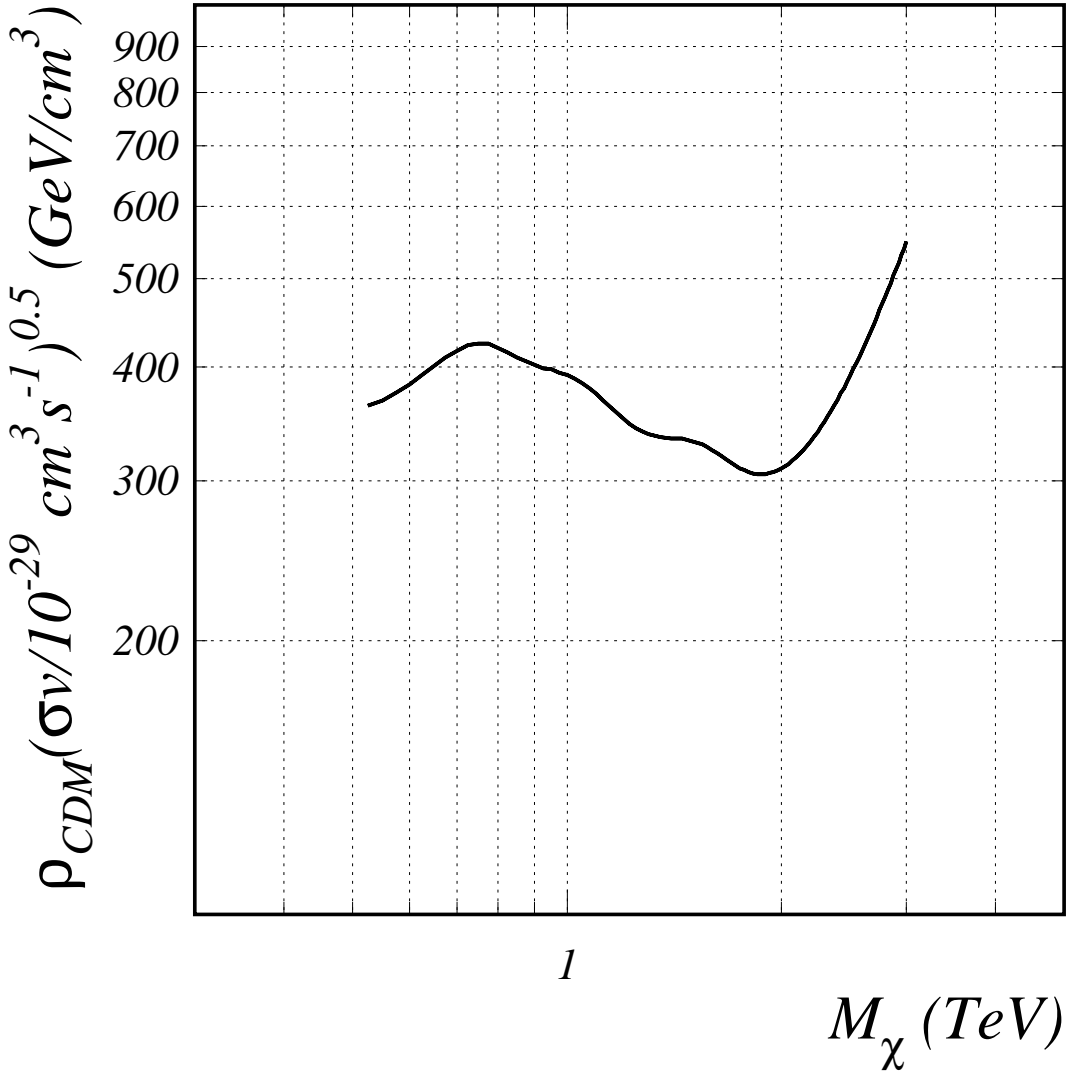}
\figcaption{
Upper limits of $\rho_{CDM}$ versus $M_\chi$. Here, 
a monochromatic gamma-ray search was carried out for
the reaction $\chi\chi\to\gamma\gamma$.
\label{fig5}
}
\end{figure}

In addition, we carried out a search for monochromatic gamma-rays
in the energy region between 0.5 and 3 TeV.
The TeV gamma-ray energy spectrum was fitted with Gaussians
under various peak-energy assumptions. 
A uniform energy resolution of 35\% was assumed.
A typical line is shown
in Fig. 2 (line-B). The 2$\sigma$-upper limits for 
the mass densities of CDM
, $\rho_{CDM}\sqrt{\frac{10^{-29}{\rm cm}^3{\rm s}^{-1}}
{<\sigma_{\gamma\gamma} v>}}$, 
were obtained ( Fig. \ref{fig5} ). 
Here we used a smaller normalization factor for $<\sigma_{\gamma\gamma} v>$
as was expected from the particle theory \citep{bergstrom}.
These upper limits
are higher than these in Fig. \ref{fig4}.

\section{Discussion}

The halo density estimated by the HI studies
is 0.015 M$_{\odot}$pc$^{-3}$ (0.57 $[{\rm GeV}{\rm cm}^{-3}]$) \citep{puche},
which is two orders lower than our upper limits.
The gamma-ray flux from NGC 253 should be explained
by the standard cosmic-ray theory.
Due to the starburst phenomena, we could not throw away a cosmic-ray
interpretation \citep{volk}.
If the cosmic-ray emission is accurately determined
from the study of multi-wavelength spectrum, this
upper limit will be greatly reduced to the error-bar level.

A scan of the nearby galaxies that are not starburst will be promising
in the search for CDM. 
Especially, the Sculptor group is an interesting
target, which is also suggested by an HI measurement \citep{puche}.

The search for gamma-rays from massive galaxies is considered to
reduce the upper limit for the galactic density of CDM.
Ten-times-heavier astronomical objects that are nearby, 
would give a sensible result.
For example, M 87 is considered to be more than ten-times 
heavier than our Galaxy \citep{baltz}.
The distance is several-times farther than NGC 253. 
The volume-distance factor, $\frac{d}{3}\cdot({\Delta \theta})^3$, is, 
therefore, an order
lower than our case (reported ${\Delta \theta}<0.127^o$ \citep{m87}).
On the other hand, the recent
observation by HEGRA reported about a ten-times fainter gamma-ray
intensity \citep{m87}.
These values should result in the same order of CDM density, 
while M87's density is considered to be larger than that of NGC 253. 
Publication of the differential
flux of the TeV gamma-ray from M87 is awaited.

Although M 31 is also massive, the visible size is larger than
the field of view of IACTs, which requires special treatments for
background subtractions \citep{m31}. 

Considering a figure of merit ($FOM$) for the CDM search, 
we had better consider total mass of galaxy, volume of halo, distance,
and visible size simultaneously. 
The CDM density may be proportional to the total mass
($M_G$) divided by the volume.
Thus, the expected gamma-ray's flux should be proportional to
the following: $FOM=M_G^2d^{-5}(\Delta \theta )^{-3}$.
Selecting those nearby galaxies that have a visible size of between
$3\times 10^{-3}$ and $10^{-2}$ radian
(favorable size for the IACT measurements), NGC 5128 (Cen-A) and
NGC 5236 (M 83) were calculated to have a bigger $FOM$ than that
of NGC 253.
Especially for Cen-A, hundred-times larger flux is expected.

We also applied the same discussion to $\omega$-Centauri
\citep{guy}.
The dark matter origin of globular clusters was
proposed by \cite{peebles}.
The distance is close and lower cosmic-ray level is expected.
The $FOM$ was calculated to be 10000 times higher than that of NGC 253.
An upper
limit which is the same order with the baryonic density
could be obtained.

Also, a high-sensitivity search in the Galactic Center is awaited
\citep{tsuchiya}.
However, to remove the model dependence and to estimate
the cosmic-ray there are keys for this case.

Compared to the accelerator experiment, only IACT measurements are
sensitive to CDM with a mass heavier than TeV.
They are complementary important to each other.

\section{Conclusion}

A constraint on the cold dark matter (CDM) was obtained
using the data of
the gamma-ray halo around the nearby starburst galaxy NGC 253.
According to this study,
upper limits for the CDM density were obtained in the mass
range between 0.5 and 50 TeV.
Although these limit is higher than the expected value, this is one of first
trials from the IACT observational side.
The IACTs have been proven to have abilities to detect it.
The presently existing IACTs are competitive devices compared with high-energy
particle accelerators. 
The nearby galaxies such as NGC 5128 (Cen-A) and NGC 5236 (M 83)
and / or globular cluster $\omega$-Centauri
will be next interesting targets.
Observational efforts for probable candidates should
be systematically continued.

\acknowledgments

We thank Prof. J. Hisano and Prof. M. Fukugita of ICRR for
various discussions.

%\end{document}

%\clearpage

\end{document}